\documentclass[aps,preprint,showpacs,showkeys,preprintnumbers,eqsecnum,amsmath,amssymb]{revtex4}

\usepackage{dcolumn}
\usepackage{amsmath,amssymb,amsthm,color}
\usepackage{epsfig}
\usepackage{graphicx}
\usepackage{amssymb}
\usepackage{amsfonts}
\usepackage{amsmath}
\usepackage{graphicx}
\usepackage{subfigure}
\usepackage{setspace}
\usepackage{graphics}
\usepackage{bm}
\usepackage{float}

\begin{document}
\baselineskip=0.8 cm
\title{Rotating non-Kerr black hole and energy extraction}

\author{
Changqing Liu$^{1,2}$, 
Songbai Chen$^{1}$, 
and Jiliang  Jing$^{1}$\footnote{Corresponding author, Email:
jljing@hunnu.edu.cn} 
}

\affiliation{1) Department of Physics, and Key Laboratory of Low Dimensional Quantum Structures \\ and Quantum Control of Ministry of Education, Hunan Normal
University, \\ Changsha, Hunan 410081, P. R. China}

\affiliation{2) Department of Physics and Information Engineering, \\
Hunan Institute of Humanities Science and Technology,\\ Loudi, Hunan
417000, P. R. China}

\begin{abstract}
\baselineskip=0.6 cm

The properties of the ergosphere and energy
extraction by Penrose process in a rotating non-Kerr black hole are investigated. It is shown that the ergosphere is sensitive to the deformation
parameter $\epsilon$ and the shape of the ergosphere
becomes thick with increase of the parameter $\epsilon$. It is of
interest to note that, comparing with the Kerr black hole, the
deformation parameter $\epsilon$ can enhance the maximum efficiency of
the energy extraction process greatly. Especially, for the case of $a>M$,
the non-Kerr metric describes a superspinning compact object and
the maximum efficiency can exceed $60\%$, while it is only $20.7\%$
for the extremal Kerr black hole.

\end{abstract}


\maketitle
\newpage
\section{Introduction}

In 4-dimensional general relativity, no-hair theorem \cite{noh}
guarantees that a neutral rotating astrophysical black hole is
uniquely described by the Kerr metric which  only possesses
two parameters, the mass $M$ and the rotational parameter $a$. For the Kerr black hole,
the fundamental limit is the bound $a\leq M$, and the central
singularity  is always behind the event horizon due to the weak
cosmic censorship conjecture \cite{WCC}.  However, the hypothesis
that the astrophysical black-hole candidates are described by the
Kerr metrics still lacks the direct evidence, and the
general relativity has been tested only for weak gravitational
fields \cite{CMW}. In the  regime of strong gravity, the general relativity
could be broken down and astrophysical black holes might not be the Kerr
black holes as predicted by the no-hair theorem \cite{FCa,TJo,CBa}.
Several parametric deviations from the Kerr metric have been
suggested to study observational signatures in both the
electromagnetic \cite{Reviews} and gravitational-wave \cite{EMRIs}
spectral that differ from the expected Kerr signals.

Recently, Johannsen and Psaltis proposed a deformed Kerr-like
metric \cite{TJo} suitable for the strong field of the no-hair
theorem, which describes a rotating black hole (we named it the non-Kerr
black hole) in an alternative theory of gravity beyond Einstein's
general relativity. The
non-Kerr black hole possesses the following novel features: there is
no restriction on the value of the rotational parameter $a$ due to
the existence of the deformation parameter $\epsilon$. Interestingly, for a positive
parameter $\epsilon$, the non-Kerr black hole becomes more prolate
than the Kerr black hole  and there are two disconnected horizons for high spin parameters, but there is no horizon when $a>M$.
Therefore, for a negative parameter $\epsilon$, the non-Kerr black
hole is more oblate than the Kerr black hole, and the horizon always
exists for an arbitrary $a$ and the topology of the horizon becomes
toroidal \cite{CBa,CBa1}. The non-Kerr metric is an ideal spacetime to carry out strong-field tests of the no-hair theorem. Therefore, a lot of effort has been dedicated to the study of the rotating non-Kerr black hole recently
\cite{FCa,TJo,CBa, CBa1,CBa2, VCa1,TJo1}. In Ref. \cite{csbchen}, we
studied the properties of the thin accretion disk in the rotating
non-Kerr spacetime and found that the presence of the deformation
parameter $\epsilon$ changes the inner border of the disk,
energy flux, conversion efficiency, radiation temperature, spectral luminosity and spectral cut-off frequency of the thin
accretion disk. Moreover, for the rapidly rotating black hole, the
effect of the deformation parameter $\epsilon$ on the physical
quantities of the thin disk becomes more distinct for the prograde
particles and more tiny for the retrograde ones. These significant
features in the mass accretion process may provide a possibility to
test gravity in the regime of the strong field  in the astronomical
observations.

The power energy for a active galactic nuclei,
X-ray binaries and quasars has always been concerned in the high energy
astrophysics. Several mechanisms (i.e. the accretion disk model
\cite{Kozfowski,shakura} and Blandford-Znajek mechanism
 \cite{Blandford}) have been proposed to interpret how to extract
energy from a black hole and the formation of the power jet.
Furthermore, the Penrose process
\cite{WCC,chandrasekhar,efficiency} also provides an important
method  to extract energy from a black hole.
The Penrose process was also extended to the five-dimensional
supergravity rotating black hole \cite{K. Prabhu}, higher
dimensional black holes and black rings \cite{Nozawa}, and
Ho\v{r}ava-Lifshitz Gravity \cite{Abdujabbarov}. In this paper, we will
investigate in detail the ergosphere of the non-Kerr black hole and
how the deformation parameter $\epsilon$ of the non-Kerr
black hole affects the negative energy state and the efficiency of the
energy extraction.

The paper is organized as follows: in Sec. II, we
review briefly the metric of the rotating non-Kerr black hole  proposed by
Johannsen and Psaltis \cite{TJo} to test gravity in the regime of the strong
field  and then analyze the ergosphere structure.
In Sec. III, we  investigate
the efficiency of the energy extraction by using the Penrose process.
Sec. IV is devoted to a brief summary.

\section{Rotating non-Kerr black hole spacetime} \label{section2222}

To test gravity in the  regime of the strong field, Johannsen and Psaltis \cite{TJo}, starting from a deformed Schwarzschild solution and applying the Newman-Janis transformation, constructed a deformed Kerr-like metric which describes a
stationary, axisymmetric, and asymptotically flat vacuum spacetime.
In the standard Boyer-Lindquist coordinates, the metric  can be expressed as \cite{TJo}
\begin{eqnarray}
ds^2=g_{tt}dt^2+g_{rr}dr^2+g_{\theta\theta}d\theta^2+g_{\phi\phi}
d\phi^2+2g_{t\phi}dtd\phi, \label{metric0}
\end{eqnarray}
with
\begin{eqnarray}
g_{tt}&=&-\bigg(1-\frac{2Mr}{\rho^2}\bigg)(1+h),\;\;\;\;\;
g_{t\phi}=-\frac{2aMr\sin^2\theta}{\rho^2}(1+h),\nonumber\\
g_{rr}&=&\frac{\rho^2(1+h)}{\Delta+a^2h\sin^2\theta},\;\;\;\;\;\;\;\;\;\;\;\;\;\;\;
g_{\theta\theta}=\rho^2,\nonumber\\
g_{\phi\phi}
&=&\sin^2\theta\bigg[\rho^2
+\frac{a^2(\rho^2+2Mr)\sin^2\theta}{\rho^2}(1+h)\bigg],
\end{eqnarray}
where
\begin{eqnarray}
\rho^2=r^2+a^2\cos^2\theta,\;\;\;\;\;\;\;\;\;\;
\Delta=r^2-2Mr+a^2,\;\;\;\;\;\;\;\;\;\;h=\frac{\epsilon M^3
r}{\rho^4}.
\end{eqnarray}
The constant $\epsilon$ is the deformation parameter.  The quantity
$\epsilon>0$ or $\epsilon<0$ corresponds to the cases in which the
compact object is more prolate or oblate than the Kerr black hole,
respectively. As $\epsilon=0$, the black hole is reduced to the
usual Kerr black hole in general relativity. The horizons of the
black hole are described by the roots of the following equation \cite{TJo}
\begin{eqnarray}
\Delta+a^2h\sin^2\theta=0.
\end{eqnarray}
Clearly, the radii of the horizons depends on $\theta$, which are
different from that in the usual Kerr case. For the case of
$\epsilon>0$, there exist two disconnected
horizons for high spin parameters, but there is no horizon
when $a>M$. However, for $\epsilon<0$  the horizons
never disappear for an arbitrary $a$ and the shape of the horizons
becomes toroidal \cite{CBa,CBa1}.

\begin{figure}
\begin{center}
\includegraphics[scale=1.0]{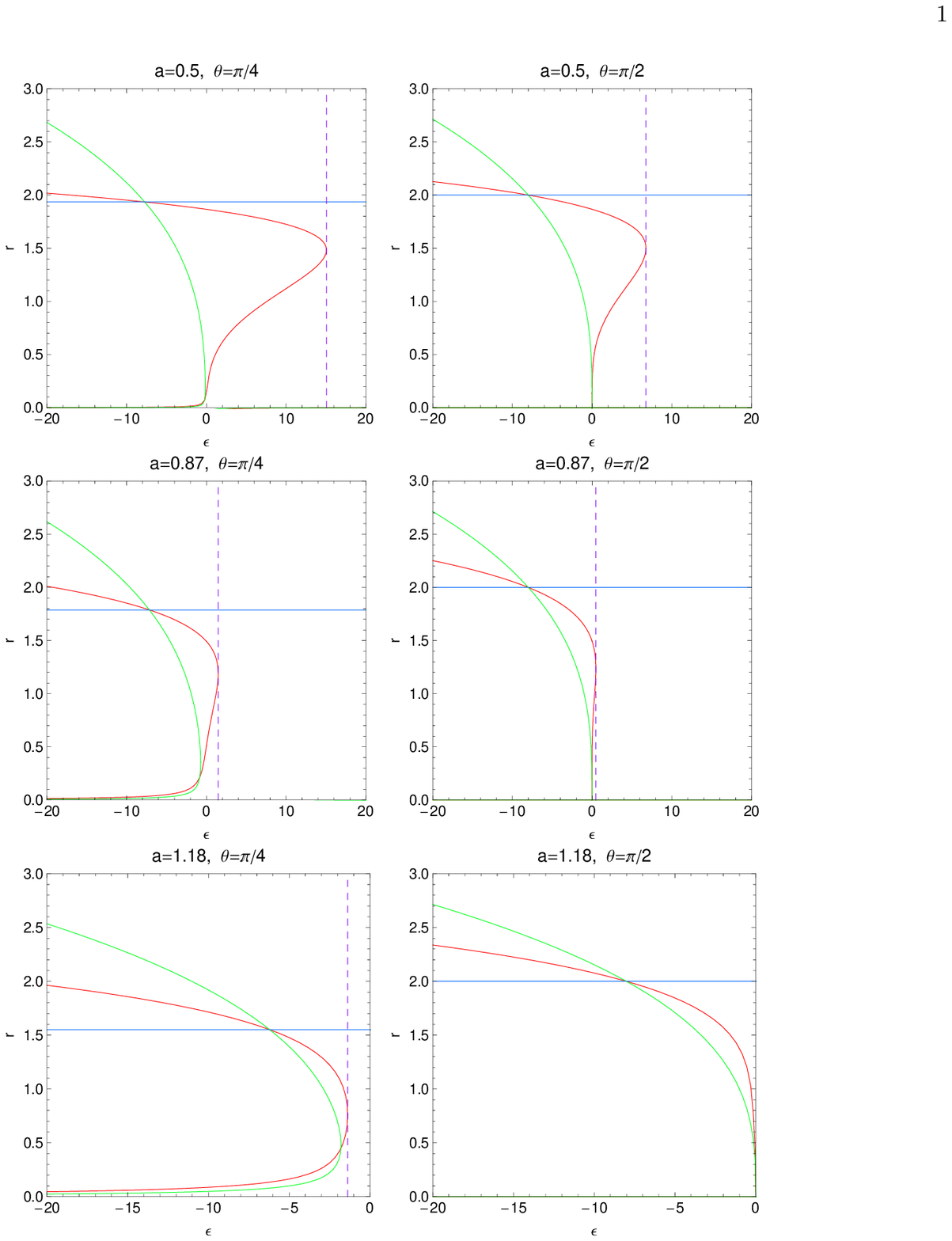} \caption{The variation of
the event horizon radius and the outer infinite redshift surface
radius with the deformation parameter $\epsilon$ of the
rotating non-Kerr black hole. The green, blue, and red lines
correspond to the surfaces defined by  $1+h=0$, $1-\frac{2Mr}{\rho^2}=0$,
and $\Delta+a^2h\sin^2\theta=0$,
respectively. The dashed line is the location of turning point of the surface
defined by $\Delta+a^2h\sin^2\theta=0$.
Here, we take  $M=1$.}\label{htt}
\end{center}
\end{figure}

The infinite redshift surface of a black hole
is defined by the roots of $g_{tt}=0$. For the non-Kerr black hole we note that $g_{tt}=0$ gives
\begin{eqnarray}
1+h=0,~~~~or~~~ 1-\frac{2Mr}{\rho^2}=0.
 \end{eqnarray}
Obviously, for the case of $\epsilon\geq 0$, the outer infinite
redshift surface is determined by $M+\sqrt{M^2-a^2cos\theta^2}$.
For  $\epsilon<0$, it seems that both of the positive root
of  $1+h=0$ and $1-\frac{2Mr}{\rho^2}=0$ are the infinite redshift
surfaces of the non-Kerr black hole. However, the determinant of the metric
$\sqrt{-g}=(1+h)\rho^2sin\theta^2$
vanishes if $1+h=0$ which shows that the surface defined by
$1+h=0$ is an intrinsic singularity. The singularity is indicated by infinite curvature
and cannot been eliminated by coordinate transformation, and it's
Kretschmann scalar $K=R^{\alpha\beta\gamma\delta}R_{\alpha\beta\gamma\delta}$ and $\sqrt{-g}$ becomes zero or infinity.
In Fig. \ref{htt}, we depict the positive root of the equations
$1+h=0$, $1-\frac{2Mr}{\rho^2}=0$, and $\Delta+a^2h\sin^2\theta=0$
with different deformation parameter $\epsilon$. The figure also shows
that the surface defined by $1+h=0$ is the intrinsic singularity,
because if we choose the surface defined by $1+h=0$ as the outer
infinite redshift surface, we find that the metric of the non-Kerr
black hole does not satisfy $g_{tt}>0$, $g_{t\phi}<0$ and $g_{rr}>0$ in the
ergosphere and the causality of the spacetime is violated. Thus
the surface defined by $1+h=0$ cannot be the infinite redshift
surface.
In Fig. \ref{htt}, there is an intersection point of
the three surfaces and the value of the abscissa for the intersection point $\epsilon_{ip}$
is $-4(M+\sqrt{M^2-a^2cos^2\theta})$. There also exists a turning
point $\epsilon_{tp}$ for the event horizon surface.
The value of the abscissa for the turning point
is located at the position $\frac{\partial\epsilon}{\partial r}=0$,
thus we get
\begin{eqnarray}\label{soluti}
10 r^4-16 M r^3+a^2(7+cos(2\theta))r^2-a^4(1+cos(2\theta))=0.
\end{eqnarray}
The maximum positive root of Eq. (\ref{soluti}) is
\begin{eqnarray}\label{solutione1}
r_{tp}&=&\frac{1}{2}\biggl(\frac{-A}{2}+\sqrt{Z_1}+
\sqrt{\frac{3A^3}{4}-2B-Z_1+\frac{-A^3+4AB}{4\sqrt{Z_1}}}\biggr)\\\nonumber
Z_1&=&\frac{A^2}{4}-\frac{2B}{3}+\frac{2^{1/3}(B^2+12C)}{3Z_2^{1/3}}+\frac{Z_2^{1/3}}{3~2^{1/3}}\\\nonumber
Z_2&=&2B^3+27A^2C-72BC+\sqrt{-(4B^2+12C)+(2B^3+27A^2C-72BC)^2}\\\nonumber
A&=&-\frac{8M}{5},~ ~~~B=\frac{a^2(7+cos(2\theta))}{10},~~~~C=-\frac{a^4(1+cos(2\theta))}{10}.
\end{eqnarray}
Using the Eqs. (\ref{soluti}) and (\ref{solutione1}), we get the
value of the abscissa for the turning point of the curve described by
$\Delta+a^2h\sin^2\theta=0$, which is
\begin{eqnarray}\label{solee}
\epsilon_{tp}=\frac{-\Delta\rho^4}{M^3a^2rsin^2\theta}\bigg|_{r=r_{tp}}.
\end{eqnarray}
The ergosphere is the region bounded by the event horizon $r_H$ and
the outer stationary limit surface $r_\infty$. A novel feature of
the ergosphere  of a black hole is that the timelike
Killing vector becomes spacelike crossing the infinite redshift
surface. An observer moving along the timelike geodesics cannot
remain static but stationary  in the ergosphere
due to the ``frame-dragging effect" \cite{CMW}.  According to the
properties of the ergosphere, the deformation parameter $\epsilon$
should satisfy following relation
\begin{eqnarray}\label{region2}
-4(M+\sqrt{M^2-a^2cos^2\theta})\leq\epsilon\leq
\frac{\Delta\rho^4}{M^3a^2rsin^2\theta}\bigg|_{r=r_{tp}}.
\end{eqnarray}
In this range of $\epsilon$, the non-Kerr black hole has a event
horizon given by largest root of $\Delta+a^2h\sin^2\theta=0$, and a
outer infinite redshift surface described by
 \begin{eqnarray}\label{region2111}
r_\infty^+=M+\sqrt{M^2-a^2cos^2\theta}.
\end{eqnarray}
When $\epsilon\geq\epsilon_{tp}$ or $\epsilon<\epsilon_{ip}$ there is no
the event horizon and the singularity becomes naked.

\begin{figure}[htp!]
\begin{center}
\includegraphics[scale=0.95]{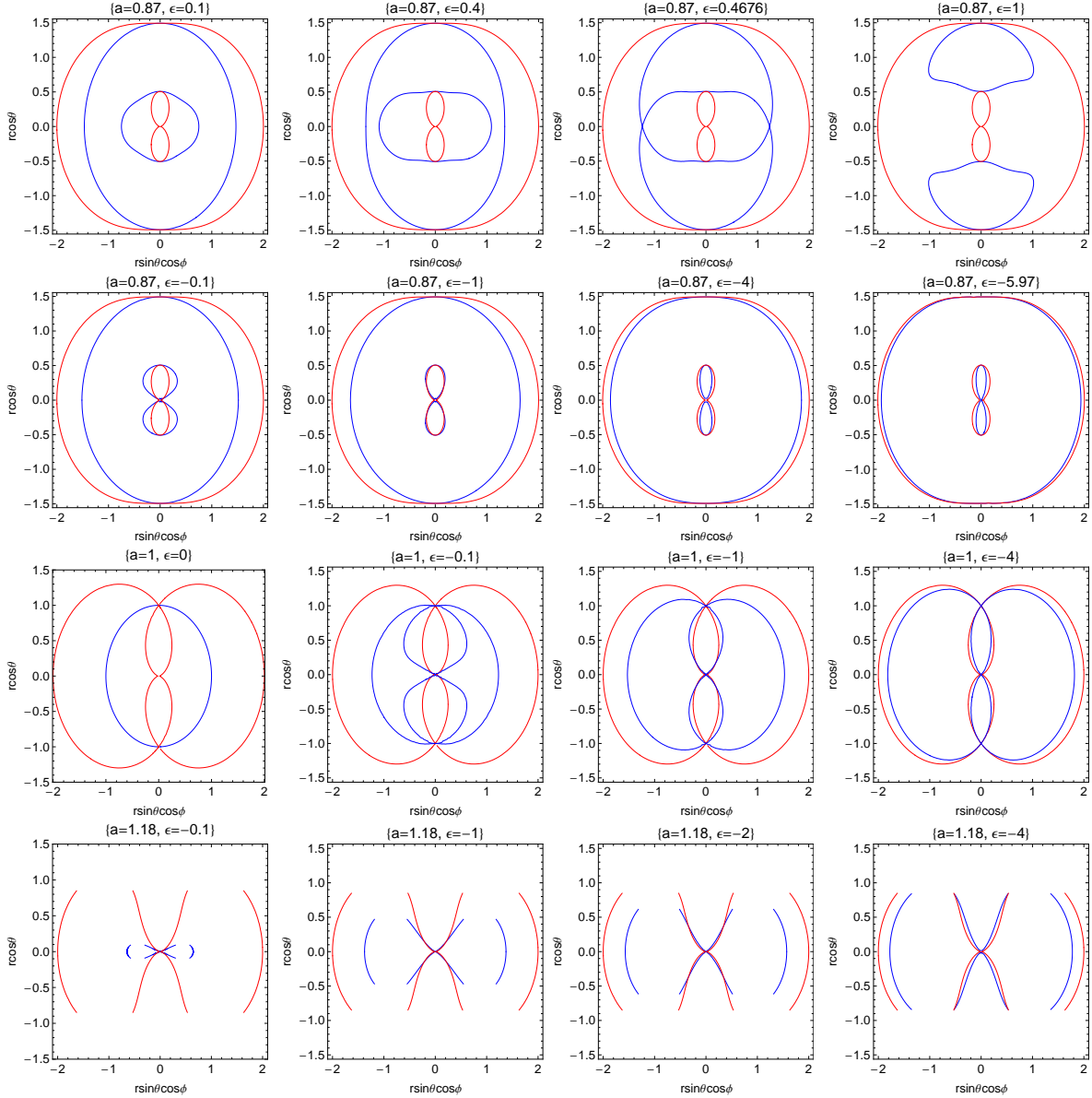}
\caption{The variation of the shape of the $xz$-plane of the
ergosphere with the deformation parameter $\epsilon$ of the
rotating non-Kerr black hole. The red and the blue lines correspond to the infinite redshift surfaces and the horizons, respectively. Here, we take
$M=1$.}\label{ergosphere11}
\end{center}
\end{figure}

How the deformation parameter $\epsilon$ affects the shape of the
ergosphere is described by Fig. \ref{ergosphere11} which shows that
the ergosphere is sensitive to the deformation parameter
$\epsilon$. For  $a<M$ (such as $a=0.87$), the non-Kerr black hole
becomes more prolate than the Kerr black hole, and the ergosphere in
the equatorial plane becomes thick as the parameter $\epsilon$
increases. It should be pointed out that, when $\epsilon$ exceed
$4.4676$, the horizons become disconnected. For the case of
$a=M$, there exist the inner and outer horizons which coincide at
the north and south poles, and the thickness of the ergosphere decreases
when the deformation parameter $\epsilon$ takes a bigger negative value. For $a>M$,
$\epsilon$ can only take a negative value, and both the horizon and
infinite redshift surface are not closed. A hole appears around the
north and south  poles in the range $\theta_{hole}\leq arccos(a/M)$.
Thus a distant observer may see the central region of this compact
object along the north or south pole. It is also interesting to
note that the overspin compact object becomes more and more thin and
looks like a disk \cite{CBa} as the deformation parameter $\epsilon$ increases.

\section{\label{enextract}Energy Extraction of the black hole by Penrose Process}

In this section, we will discuss the Penrose process
\cite{WCC,chandrasekhar,efficiency,Nozawa}, by which we can extract
a rotational energy from the non-Kerr black hole. We focus our
attention on how the deformation parameter $\epsilon$ of the
non-Kerr black hole affects the negative energy state and the
efficiency of the energy extraction.

\subsection{\label{negative energy state}The negative energy state of the  Penrose process}

 Let us now consider the trajectory of a test particle with the mass $\mu$
on the equatorial plane. With the help of the timelike Killing vector $\xi^{a}=\left(\frac{\partial}{\partial t}\right)^{a}$
and spacelike one $\psi^{a}=\left(\frac{\partial}{\partial \phi}\right)^{a}$, we have
the following conserved quantities along a timelike geodesics on the equatorial plane
\begin{eqnarray}
E&=&-g_{ab}\xi^{a}u^{b}=(1+\frac{\epsilon~M^3}{r^3})(1-\frac{2M}{r})u^{t}+
(1+\frac{\epsilon~M^3}{r^3})\frac{2aM}{r}u^{\phi},
\label{eq:e}\\
L&=&g_{ab}\psi^{a}u^{b}=-(1+\frac{\epsilon~M^3}{r^3})\frac{2aM}{r}u^{t}+
(r^2+a^2+\frac{a^2M^3(a^2+r^2)\epsilon}{r^5}+\frac{2a^2M}{r})u^{\phi},
\label{eq:L}
\end{eqnarray}
where $u^b$ is the four-velocity defined by $u^b=\frac{dx^b}{d\tau}$, $\tau$ is the proper time for the spacetime. In Eq. (\ref{eq:e}) or (\ref{eq:L}), the first equality is the basic definition of the energy or angular momentum \cite{CBa2}, and the second equality describes the energy or angular momentum of the non-Kerr black hole. Moreover, we can introduce a new
conserved parameter
\begin{eqnarray}\label{eq:phidot111}
\kappa=g_{ab}u^{a}u^{b},
\end{eqnarray}
whose values are given
by $\kappa=-1, 0, 1$ corresponding to the timelike,
null and spacelike geodesics, respectively.

From Eqs. (\ref{eq:e}), (\ref{eq:L}) and (\ref{eq:phidot111}) we can
easily obtain the equation of motion
\begin{eqnarray}\label{hamjameqq}
\alpha E^2 -2 \beta E +\gamma = 0,
\end{eqnarray}
with
\begin{eqnarray}
& \alpha=& (r^2+a^2+\frac{a^2M^3(a^2+r^2)\epsilon}{r^5}+\frac{2a^2M}{r})
\Gamma^{-1} \ , \\
&\beta=&L(1+\frac{\epsilon~M^3}{r^3})\frac{2aM}{r}\Gamma^{-1}\ , \\
& \gamma =& -L^2(1+\frac{\epsilon~M^3}{r^3})(1-\frac{2M}{r}) \Gamma^{-1}-\frac{r^2(r^3+\epsilon~M^3)}{ r^3\Delta+a^2\epsilon~M^3}
(u^r)^2-\mu^2,
\end{eqnarray}
where
\begin{eqnarray}
& \Gamma =&(1+\frac{\epsilon~M^3}{r^3})(\Delta+a^2\frac{\epsilon~M^3}{r^3}).
\end{eqnarray}

From  Eq. (\ref{hamjameqq}), we can  obtain the energy  $E$
\begin{eqnarray}
E=\frac{\beta+\sqrt{\beta^2-\alpha\gamma}}{\alpha},
\end{eqnarray}
where we only choose $+\sqrt{\beta^2-\alpha\gamma}$ to ensure that
the 4-momentum of the particle is future directed. In the Penrose
process, the orbit of the particle with negative energy in the
ergosphere is the key to extract energy from the non-Kerr black
hole. When a particle enters the ergosphere, the timelike Killing
vector becomes spacelike one, thus the energy of the particle
$E=-g_{ab}\xi^{a}u^{b}$ becomes negative. The orbit of the particle
with the negative energy $E$ must satisfy the conditions:
$\alpha>0$, $\beta<0$ and $\gamma>0$, which can be achieved only if
$La<0$.
\begin{figure}[htp!]
\begin{center}
\includegraphics[width=8cm]{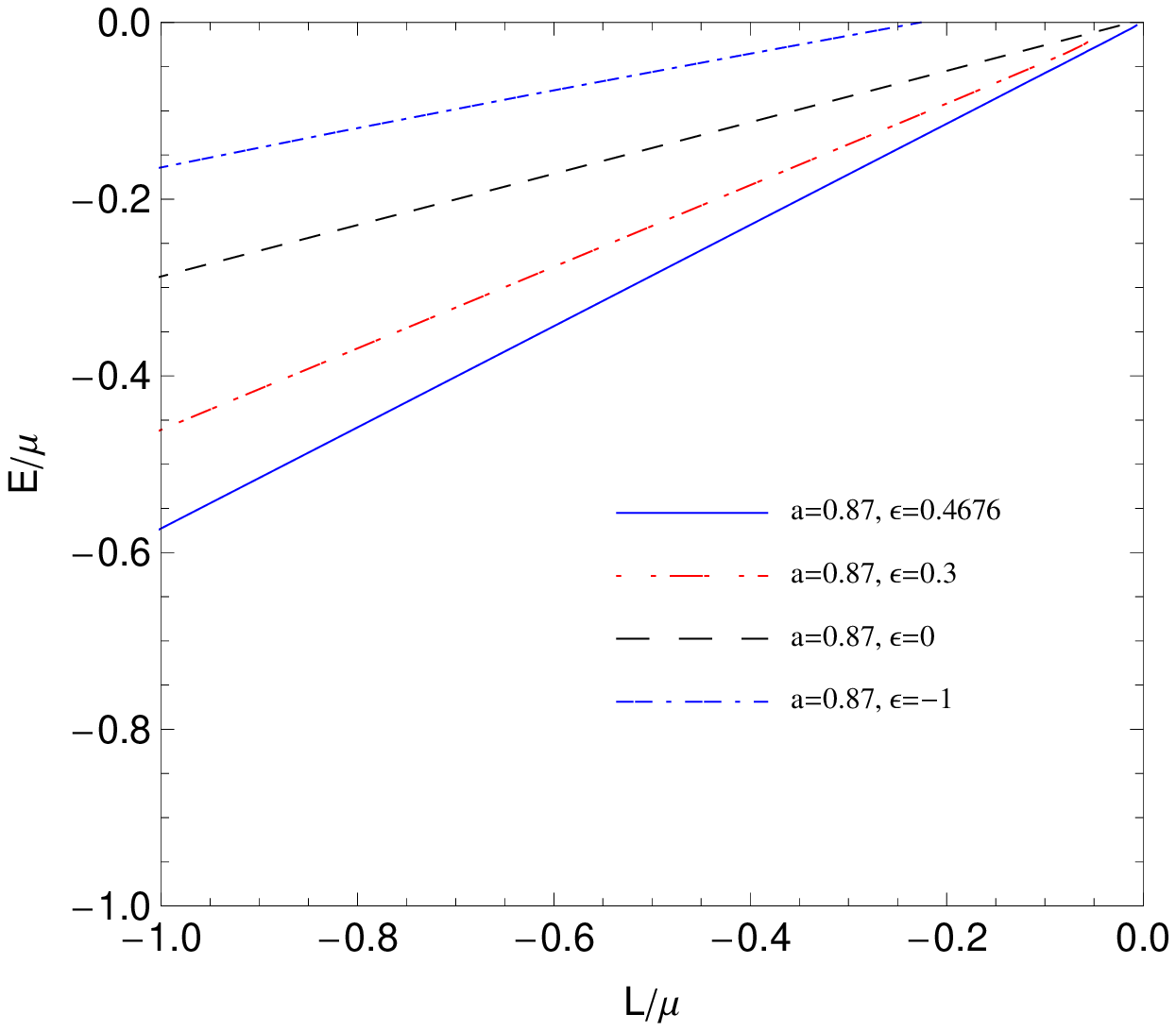}\includegraphics[width=8cm]{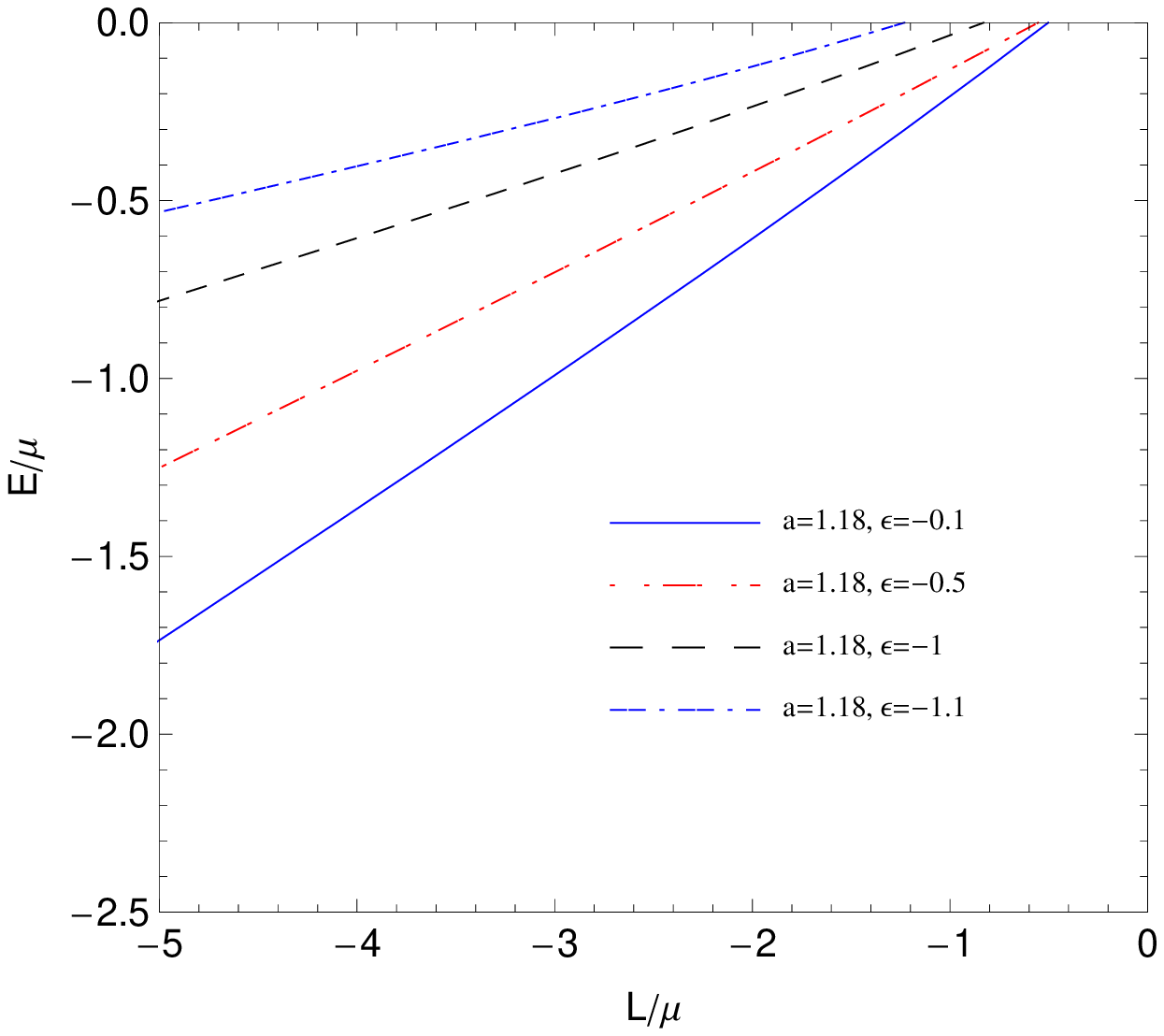}
\caption{The negative  energy state $E$ allowed for  the angular momentum $L$ and rest mass
$\mu$ of the particles for the different deformation parameter $\epsilon$ at a given location  near the event horizon inside the ergosphere.}  \label{negativeff}
\end{center}
\end{figure}
In Fig. \ref{negativeff}, we describe the negative energy state $E$
for the different deformation parameter $\epsilon$ at a given
location  near the event horizon inside the ergosphere. We find that
the negative energy $E$ increases as the deformation parameter
$\epsilon$ increases for both the cases  $a>M$ and
$a<M$.

According to the Penrose process, the mass of the black hole will
change a quantity $\delta M = E$ as a negative particle is injected
into the central black hole. Clearly $\delta M$ can be  made as
large as we wished by increasing the mass $\mu$ of the injected
particle. However, there is a lower limit on $\delta M$ which could
be added to the black hole corresponding to $\mu=0$ and
$u^r=0$~\cite{mtw}. Evaluating all of the required quantities at the
horizon $r_{H}$, we can get the lower limit
\begin{eqnarray}\label{minenergy}
E_{\rm min}=\frac{L(1+\frac{\epsilon~M^3}{r_H^3})\frac{2aM}{r_H}} {r_H^2+a^2+\frac{a^2M^3(a^2+r_H^2)\epsilon}{r_H^5}+\frac{2a^2M}{r_H}}
 .
\end{eqnarray}
From the Eq. (\ref{minenergy}) we can conclude that, in order to
extract energy from the black hole, the angular momentum of the
injected  particle must satisfy $L<0$, and the deformation
parameter $\epsilon$ influences the value of $E_{\rm min}$.

\subsection{\label{Efficiency1}Efficiency of the energy  extraction process}

The efficiency of the energy extraction process is one of the most
important questions in the  energetics of the black hole. Thus, it is
interesting to study how the deformation parameter $\epsilon$ affects
the efficiency of the Penrose process \cite{WCC,chandrasekhar,
efficiency} for the non-Kerr black hole. To calculate the maximum
efficiency of the energy extraction, we take
the radial velocity to be zero. Let $U_i$ denotes the four-velocity
of the $i$th particle of the locally nonrotating frame
observer  \cite{CMW} at a given radius $r$, which can be expressed as
\begin{eqnarray}\label{four velocity1}
U_i=u^t(1,0,0,\Omega_i),
\end{eqnarray}
with
\begin{eqnarray}\label{four velocity2}
u^t&=&-\frac{E}{X_i},~~~~X_i=-(g_{tt}+g_{t\phi}\Omega_i),\\\nonumber
\Omega_{i}&=&\frac{-g_{t\phi}(1+g_{tt})+\sqrt{(1+g_{tt})
(g_{t\phi}^2-g_{tt}g_{\phi\phi})}}
{g_{\phi \phi}+g_{t\phi}^2}\,,
\end{eqnarray}
where $\Omega_i$ is the angular velocity of the particle $i$ with respect to an asymptotic infinity observer.
In the ergosphere, $\Omega_i$ takes the value in the range
of $\Omega_-<\Omega<\Omega_+$, where
\begin{equation}
 \Omega_{\pm}=\frac{-g_{t\phi}\pm \sqrt{g_{t\phi}^2-g_{tt}g_{\phi\phi}}}{g_{\phi\phi}}.
\end{equation}
In the Penrose process, an incident particle $1$
with  the rest mass $\mu_1=1$, i.e. $E_1=1$, splits into the particle $2$ absorbed by the black hole and the particle $3$ escaping to infinity.
According to the conservational laws of the energy and angular momentum, we get
 \begin{eqnarray}\label{kk3}
U_1=\mu_2U_2+\mu_3U_3.
\label{angular_cons}\end{eqnarray}
The efficiency of the Penrose process is defined as
\begin{eqnarray}
 \eta =\frac{\mu_3E_{3}-E_{1}}{E_{1}}=\mu_3E_3-1.
\end{eqnarray}
The maximum efficiency $\eta$ can be obtained by the choice
of $\mu_2U_2$ and $\mu_3U_3$ \cite{efficiency}
\begin{eqnarray}\label{kk2}
\mu_2U_2&=&k_2(1,0,0,\Omega_-)\\\nonumber
\mu_3U_3&=&k_3(1,0,0,\Omega_+),
\end{eqnarray}
where $k_2$ and $k_3$ are constants to be determined.
With the help of the Eqs. (\ref{four velocity1}), (\ref{four velocity2}), (\ref{kk3}) and (\ref{kk2}),
we find
\begin{eqnarray}
\eta &=&
\frac{(\Omega _{1}-\Omega _{-})(g_{tt}+g_{t\phi}\Omega _{+})}{(\Omega _{+}-\Omega _{-})(g_{tt}+g_{t\phi}\Omega _{1})}-1.
\label{eq:efficiency11111}
\end{eqnarray}
When the incident particle $1$ splits at the horizon $r_H$,
we thus obtain the maximum efficiency
\begin{eqnarray}
\eta_{max}=\frac{\sqrt{1+g_{tt}}-1}{2}\bigg|_{r=r_H}.
\label{eq:efficiencymax}
\end{eqnarray}
\begin{figure}[htp!]
\begin{center}
\includegraphics[width=8cm]{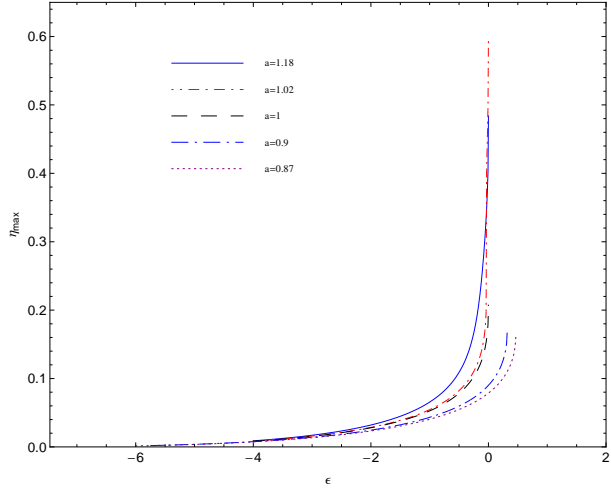}
\caption{The Variation of the maximum efficiency of the energy
extraction process with the deformation parameter $\epsilon$ of
 rotating non-Kerr black hole. Here we take $M=1$.}\label{effffffffffff}
\end{center}
\end{figure}

Now, we would like to analyze the effects of $\epsilon$ on the
efficiency of the rapidly rotating non-Kerr black hole. We calculate
the maximum efficiency with the numerical method and present the
variation of the maximum efficiency in the energy extraction process
with the deformation parameter $\epsilon$, which takes the value in the
range (\ref{region2}) to guarantee that the non-Kerr black hole  has
the connected event horizon for a fixed $a$
in Fig. \ref{effffffffffff}. We also present the effects of the parameter $\epsilon$ on the
maximum efficiency in Table. \ref{table1} and  \ref{table2}.\begin{table}[htp11]
\begin{center}
\caption{The maximum efficiency $\eta_{\rm max}$ of energy
extraction process in the non-Kerr black hole depends on  the
 parameter $\epsilon$  for  $a\leq M$. Here we set M=1.}
\begin{tabular}{c c c c c c c c}
  \hline
  \hline
   & $a$=0.2 & $a$=0.4 & $a$=0.6 & $a$=0.8 & $a$=0.9  & $a$=0.99 & $a$=1\\
  \hline $\epsilon$=0\;\; &  0.25\%  & 1.0\% & 2.7\% & 5.9\% & 9.01\% &16.2\% & 20.7\%\\
 $\epsilon$=0.01  & 0.255\% & 1.082\% & 2.710\% & 5.940\% &9.1\% & 17.1\% &  \\
 $\epsilon$=0.02  & 0.256\% & 1.083\% & 2.73\% & 5.975\% & 9.200\% & 19.025\% &  \\
 $\epsilon$=0.2  & 0.268\% & 1.142\% & 2.915\% & 6.71\% &11.58\% &  &  \\
 $\epsilon$=0.3  & 0.275\% & 1.175\% & 3.026\% & 7.187\% & 14.594\% &  & \\
 $\epsilon$=0.4  & 0.282\% & 1.209\% & 3.242\% & 8.442\% &  &  & \\
  \hline\label{table1}
\end{tabular}
\end{center}
\end{table}

\begin{table}[htp!]
\begin{center}
\caption{The maximum efficiency $\eta_{\rm max}$ of energy
extraction process in the non-Kerr black hole depends on  the
parameter $\epsilon$ for  $a>M$. Here we set M=1.}
\begin{tabular}{c c c c c c c c}
  \hline
  \hline
   & $a$=1.001 & $a$=1.01   & $a$=1.1 & $a$=1.15 & $a$=1.2\\
  \hline $\epsilon$=-0.00001\;\; & 60.739\%  & 59.954\% &  52.885\% &49.487\%\% &46.406\% \\
 $\epsilon$=-0.0001  & 59.577\% & 58.806\% & 51.859\% & 48.520\% & 45.492\%  \\
 $\epsilon$=-0.001  & 56.922\% & 56.186\% & 49.547\% & 46.353\% &43.455\%  \\
 $\epsilon$=-0.01  & 50.088\% & 49.492\% & 43.945\% & 41.201\% & 38.683\%  \\
 $\epsilon$=-0.1  & 13.136\% & 13.844\% & 24.878\% & 26.231\% & 25.901\% \\
 $\epsilon$=-1  & 5.278\% & 5.360\% & 6.087\% & 6.420\% & 6.638\% \\
  \hline\label{table2}
\end{tabular}
\end{center}
\end{table}
Obviously, it is shown that the maximum efficiency of the Penrose
process can be enhanced as  the parameter $\epsilon$ increases.
It is interesting to note that, for  $a>M$, the non-Kerr metric
describes a superspinning black hole \cite{TJo,C. Bambi1}, the
maximum efficiency can exceed $60\%$, while it  is only $20.7\%$ for the
extremal Kerr black hole.  This result is reasonable, because in the
Sec. \ref{section2222}, we investigate the ergosphere and
find that the maximum thickness of the ergosphere on the equatorial
plane  increases as the deformation parameter
$\epsilon$ increases. Especially, for  $a=1.18$ and $\epsilon=-0.1$, the
thickness of the ergosphere on the equatorial plane is much thicker
than that of the extremal Kerr black hole ($a=1$, $\epsilon=0$)  in Fig.
\ref{ergosphere11}.
  If the values of $a$ and $\epsilon$ change gradually, by accretion
 or any other process, from $a<M$ to $a>M$ and from $\epsilon>0$ to
  $\epsilon<0$ we should expect a continuous change
  of the energy extraction intuitively. Why is it not so?
  From the Eq.
(\ref{eq:efficiencymax}), we find that the maximum efficiency
($\eta_{max}=\frac{\sqrt{1+g_{tt}}-1}{2}\big|_{r=r_H}$) is related
to the event horizon $r_H$ of the non-Kerr black hole. In Fig. 5, by taking $\epsilon=-0.001$ and considering the values of $a$ changes gradually
from $a=1$ to $1.001$, we find the variation of the event horizon $r_H$ is not continuous. Therefore, we can not obtain a continuous change of the energy extraction for the process by taking $a$ from $a<M$ to $a>M$ (or from $\epsilon>0$ to
  $\epsilon<0$).
\begin{figure}[htp11]
\begin{center}
\includegraphics[width=8cm]{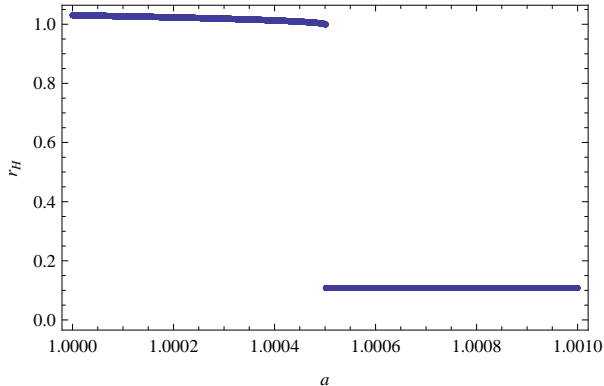}
\caption{The variation of the event horizon
 with the rotating parameter $a$ of a
 rotating non-Kerr black hole with $M=1$ and $\epsilon=-0.001$ .}\label{rhaaa}
\end{center}
\end{figure}

\section{summary}

In this paper, we present a detail analysis of the properties of the
ergosphere in the rotating non-Kerr black hole proposed
recently by Johannsen and Psaltis \cite{TJo} to test gravity in
the regime of the strong field in the future astronomical observations.
We now summarize our results as follows:
 (1) We present the restricted conditions on the deformation
parameter $\epsilon$ to guarantee that the non-Kerr black hole has
the connected  horizons (see Eq. \ref{region2} and Fig. \ref{htt}).
(2) We show that the ergosphere is sensitive to the
deformation parameter $\epsilon$ (see Fig. \ref{ergosphere11}) and the shape
of the ergosphere becomes thick with increase of the deformation
parameter $\epsilon$. (3) We find that, comparing with the Kerr black hole, the deformation parameter
$\epsilon$ not only enlarges the negative energy $E$ (see Fig. \ref{negativeff})
 but also enhances the maximum
efficiency of the energy extraction process (see tables \ref{table1}-\ref{table2} and Fig. \ref{effffffffffff}).  Moreover, the maximum
efficiency can exceed $60\%$ for the non-Kerr compact objects with
$a>M$. The influence of
the deformation parameter $\epsilon$ on the maximum efficiency
presents a good theoretical opportunity to distinguish the non-Kerr
black hole from the Kerr one and to test whether or not the
current black-hole candidates are the black holes predicted by Einstein's
general relativity. However, we think such a test is not possible at present.

\section{Acknowledgments}

This work was supported by the National Natural Science  Foundation
of China under Grant No. 11175065, 10935013; the National Basic
Research of China under Grant No. 2010CB833004; the SRFDP under Grant
No. 20114306110003; PCSIRT, No. IRT0964;
the Hunan Provincial Natural Science Foundation of China under Grant
No 11JJ7001; Hunan Provincial Innovation Foundation For Postgraduate
under Grant No CX2011B185; and Construct Program of the National
Key Discipline.

\end{document}